Title: Gravitational tempering in colloidal epitaxy to reduce defects further

Authors: Atsushi Mori[*,1], Yoshihisa Suzuki[1], Masahide Sato[2]

Affiliation: [1]Institute of Technology and Science, The University of Tokushima, [2]Information Media Center, Kanazawa University

Abstract: Less-defective colloidal crystals can be used as photonic crystals. To this end, colloidal epitaxy was proposed in 1997 as a method to reduce the stacking defects in the colloidal crystals. In this method, face-centered cubic (fcc) (001) stacking is forced by a template. In fcc (001) stacking, in contract to fcc {111} stacking, the stacking sequence is unique and thus the stacking fault can be avoided. Additionally, in 1997, an effect of gravity that reduces the stacking disorder in hard-sphere (HS) colloidal crystals was found. Recently, we have proposed a gravitational tempering method based on a result of Monte Carlo (MC) simulations using the HS model; after a colloidal crystal is grown in a relatively strong gravitational field, the defects can be reduced by decreasing the gravity strength and maintain for a period of time. Here, we demonstrate this method using MC simulations with a programed gravitation. The dramatic disappearance of defect structures is observed. Gravitational tempering can complement gravitational annealing; some defect structures that accidentally remain after gravitational annealing (keeping the colloidal crystal under gravity of a considerable constant strength) can be erased.

[*] Corresponding author

    Full name: Atsushi Mori

    Address: Institute of Technology and Science, The University of Tokushima, 2-1 Minamijosanjima, Tokushima 770-8506, Japan

    Phone: +81-88-656-9417

    Fax: +81-88-656-9435

    E-mail: atsushimori@tokushima-u.ac.jp

Controlling defects in crystals is inevitable for functionalization of materials. Because electronic and photonic band structures are calculated for perfect crystals, the calculated properties are expected for crystals with a low amount of defects. In turn, certain controlled defects may exhibit a special function. For example, certain line defects in a photonic crystal can work as optical waveguides. In this paper, we focus on colloidal crystals, where the entities of the crystals can be directly manipulated because colloidal particles are larger than atoms by several orders of magnitude. However, it is too time-consuming to manipulate the colloidal particles in the colloidal crystals. Therefore, the methods that are based on spontaneous processes such as phase transition kinetics, defect dynamics, and so on are expected.

The crystalline phase in the hard-sphere (HS) system was discovered in 1957[1,2]. The phase behaviour of the HS system is governed by only the particle number density; the coexistence region of the fluid and crystalline phases is independent of the temperature[3,4]. As the density increases, the HS system crystallises. Accordingly, the HS suspensions and the charge-stabilised colloids, which have a repulsive screened Coulombic interaction, crystallize by sedimentation[5].

In the scope of photonic crystal applications of the colloidal crystals, defect reduction in the colloidal crystals is required. The use of a patterned substrate as a template (the colloidal epitaxy)[6] is promising. In addition, by using a square pattern as a template, one can force the crystal growth in the face-centered cubic (fcc) <001> direction[7-9]. Stacking faults can be avoided because the stacking sequence in this stacking is unique. In contrast, the stacking disorders naturally occur in fcc {111} stacking. Furthermore, in a series of studies[10-18], we focused on the effect of gravity that reduces the stacking disorders in colloidal crystals, which was firstly found by Zhu et al.[19] and similar results[20,21] supporting this trend were obtained later.

We found a glide mechanism of a Shockley partial dislocation that can shrink an intrinsic stacking fault in Monte Carlo (MC) simulations[13]. We had a support to this mechanism by an elastic field calculation[14,22] in a coherent growth[12]. The shortcoming of these simulations was that fcc (001) stacking was driven by a stress from a small simulation box. The first reason for our successive simulations[15-17] was to resolve this artefact. We have successfully performed fcc (001) stacking in the MC simulations of the colloidal epitaxy on a square pattern. The second reason was to reproduce the disappearance of the stacking faults using a realistic driving force for fcc (001) stacking. The separation of the lattice line projections has vanished, which can be explained as the disappearance of a stacking fault. If a stacking fault runs across a lattice line, this lattice line splits into two lines, which are separated by a Burgers vector of the Shockley partial dislocation. In addition, the avoidance of polycrystallisation[16,17], which is another advantage of the colloidal epitaxy, was observed. In the colloidal epitaxy, a crystal nucleation on the patterned substrate and a successive upward growth will dominate over the crystal nucleation inside a region apart from the bottom if the gravity strength is not too large. If the gravitational strength is sufficiently large, the nucleation inside occurs prior to the propagation of the front of the crystal that grows from the bottom. The third reason was that we expected to observe some phenomena that were not observed when the system was small. The expected phenomena include the finding of triangular-shaped defects[16,17].

In a previous study[18], we proposed a new method, which we called gravitational tempering, to reduce the defects in colloidal crystals further. In addition to gravitational annealing (maintaining the colloidal crystal in a gravitational field of a certain strength for a considerable period of time)[20,21,23-28], gravitational tempering is in effect. If defects remain in a colloidal crystal grown in a gravitational field of certain strength, some of those defects can be erased by maintaining the system in moderately low gravity. In this paper, we demonstrate gravitational tempering using MC simulations with a programed gravitation.

In a previous study[15], we controlled the dimensionless quantity $g^* \equiv mg\sigma/k_BT$ (which is called the gravitational number and is equal to the reduced inverse gravitational length $\sigma/l_g$ with $l_g \equiv k_BT/mg$) step-wise as previously proposed[11] to avoid polycrystallisation (the HS system polycrystallized by a sudden switching-on of gravity in a case of the flat bottom wall in a previous simulation study[10]). Here, $m$ is the mass of a particle, $g$ is the acceleration due to gravity, $\sigma$ is the HS diameter, and $k_BT$ is the temperature multiplied by Boltzmann's constant. In Refs. 16 and 17, gravity was suddenly switched-on. In the present study, gravity was applied to an initial random configuration and maintained for $2 \times 10^7$ Monte Carlo cycles (MCCs). Here, one MCC is defined to include $N$ MC particle position moves where $N$ is the number of particles. After that, gravity was decreased and maintained for $2 \times 10^7$ MCCs. Finally, gravity was reset to the initial value. In this article, we report the results for $g^*$ = 1.6-1.4-1.6 gravitation program ($g^*$ = 1.6 for the initial and final $2 \times 10^7$ MCCs and $g^*$ = 1.4 for the mid $2 \times 10^7$ MCCs). We have inspired by a previous study[18], where the amount of defects was smaller for $g^*$=1.4 than for $g^*$ = 1.6. The runs with five different series of random numbers were examined. Runs of a program of $g^*$ = 2.0-1.4-2.0 are in progress. Note that $g^*$ can be directly controlled by changing the rotation velocity in centrifugal sedimentation methods[23-28].

We have performed the conventional MC simulations in this series of studies. The system size was fixed, and the MC moves of the individual particle positions were attempted. After each move, if an overlap of particles occurs, the new configuration is rejected. The standard Metropolis method based on gravitational energy was employed. The maximum displacement of the particles in the present runs was 0.06σ. For future studies, note that their transformation into kinetic MC simulations is simply accomplished by scaling the time[29,30]: the MCCs are divided by the acceptance ratio and measured in the unit of $\sigma^2/D_0$, where $D_0$ is the diffusion coefficient in the infinite dilution. However, this procedure is unnecessary for the present purpose. We do not aim to evaluate the rate of structural change such as the shrink of the stacking fault, and the activation barrier is not evaluated using the rate of structural change. In the future, we will evaluate the rates of motion of some types of defects by transforming the conventional MC simulations into kinetic simulations. We would like to add more a remark concerning the real colloidal systems. In the real colloidal systems the diameters of the particles are not identical. The polydispersity in the size of the colloidal particles should be taken into account in comparison to the real system. In this respect, the present study using an idealized model will provide a new qualitative concept. We note that the ordering is, in general, suffered from the polydispersity.

The system size of the present simulation was $L_x = L_y = 25.092688\sigma$ and $L_z = 1000.00000\sigma$ (we essentially borrow the expressions of double-precision floating points in this paragraph for clarity – see note 18 of ref. 22). $N = 26624$ HSs were placed between the well-separated top (at $z = L_z$) and bottom (at $z = 0$) walls. The periodic boundary condition was imposed in the horizontal direction. The pattern on the bottom wall was identical to that on the previous one[15]. The groove width was $0.707106781\sigma$, and the side-to-side separation between two adjacent grooves was $0.33842118\sigma$. The diagonal length of the intersection of the longitudinal and transverse grooves was $0.9999999997\sigma$. Thus, an HS only fitted to but did not fall into an intersection.

The evolution of the centers of gravity ($z_G \equiv (1/N)\sum_{i=1}^{N} z_i$, where $i$ is the identification number of particles) for the first run is shown in Fig. 1a. To plot the evolution of the centers of gravity, we took the moving-block averages of $z_G$ for 10000 MCCs. Disappearance of a defect structure accompanying particle number deficiency can be detected by monitoring the center of gravity. Also, we plot the evolution of fcc- and hcp-like, and other crystalline particles in region $z \leq 20\sigma$ in Fig. 1a. Classification of crystalline particles was based on $Q_l$ and $w_l$ order parameters[31,32] as done recently[33,34]. The center of gravity rose at the instant when $g^*$ was decreased to 1.4, and simultaneously the number of fcc-like particles decreased and the numbers of hcp-like and other crystalline particles increased. A sudden sink of the center of gravity when $g^* = 1.4$ is observed. Also, corresponding increase of the number of fcc-like particles and decrease of those of hcp-like and other crystalline particles are seen. Comparing the snapshots (*xz*- and *yz*- projections) before and after this instant (Figs. 1c and d), we find that this phenomenon is caused by the disappearance of an extended triangular-shaped defect around $(x/\sigma, y/\sigma, z/\sigma) = (-10, 3, 15)$. We can also confirm the defect disappearance in an animation ([Supplementary Video 1](Supplementary Video 1)). A snapshot at the $2 \times 10^7$th MCC, which is at the end of the initial $g^*=1.6$ duration, is shown in Fig. 1b. In this figure, we confirm a number of defects. In the animation, we observe an enlargement of the defective region in the upper portion ($z/\sigma \gtrsim 16$) according to the decrease of $g^*$. In addition, the defect configuration is nearly identical to that at the $2 \times 10^7$th MCC, except for the disappearance of a defect at $(x/\sigma, y/\sigma, z/\sigma) \approx (-7, 4, 8)$. This disappearance of a defect accompanies no significant sink of the center of gravity. Lastly, let us check the effect of returning of $g^*$ to 1.6. In the animation, we see a shrink of the defective region in the upper portion; the fluctuations of the projected positions of two lattice planes seemingly disappeared. By looking at the animation, we find that the boundary between the defective and the regular regions moves up and down as if they are breathing for $2.4$-$4 \times 10^7$ MCCs ($g^* = 1.4$), but the motion is suppressed for $4$-$6 \times 10^7$ MCCs ($g^*=1.6$). Interestingly, whereas the center of gravity sank at the instant when $g^*$ was returned to 1.6, any kinks are not observed on the lines of evolution of the numbers of crystalline particles. It means that this sink was involved in ordering, including reduction of defects, in the region $z > 20\sigma$.

With the second random numbers, the defect disappearance because of gravitational tempering did not occur. The amount of defects in this run was already small compared to that in Fig. 1b. The situations regarding the third and the fourth random numbers are nearly identical. In these runs, no decrease of $z_G$ was observed and no dramatic erases of the defect structures occurred. We can say that if the amount of defect is already small, the defects that can undergo dramatic erasing because of gravitational tempering such as performed in moderate gravity strength in this simulation no longer exist.

Let us examine the results of the run with the fifth random numbers. The evolution of $z_G$ along with the evolution of the numbers of crystalline particles is shown in Fig. 2a, where no sinks of the centre of gravity and no kinks on the lines of the evolution of the numbers of the crystalline particle after $2\times10^7$th MCC are observed. However, the extent of the defect structure reduced, which was less dramatic than that in Fig. 1, but definite. The defect configuration is nearly identical at $2\times10^7$th MCC (the end of the $g^* = 1.6$ duration) and $2.5\times10^7$th MCC (in the middle of the $g^* = 1.4$ duration) (Figs. 3b and c), and it significantly changes at $2.5\times$ and $2.6\times10^7$th MCCs (Figs. 3c and d). The four-layer-thick defect structure diminished to become two-layer thick during the $2.5$-$2.6\times10^7$th MCC period. A change in the defect configuration is also confirmed in an animation ([Supplementary Video 2](#)). The reason why this change accompanies no significant sink of the center of gravity is that only the motion of the deficiencies from the bottom of the plane defects (lines in appearance in the projections) to their tops were related to the change in the center of gravity. This phenomenon was not detected also in the evolution of the numbers of the crystalline particles, despite that disappearance of the stacking disorders did make changes between fcc- and hcp-like particles. We speculate that the changes in the numbers of the crystalline particles were small relative to the total numbers.

When the gravitational tightness at which the colloidal crystal was grown is alleviated, the defects that accidentally remained in this gravitational annealing process can move. This is the scenario of erasing defects because of gravitational tempering. In addition, we speculate that the enlargement of fluctuation of the lattice lines according to the alleviation of the tightness must play an important role. By wandering of the lattice line, vacancies and particles can interchange with each other. There is, however, a trade-off; as mentioned in an introductory part, the thermodynamic driving force for the colloidal crystallization decreases if the strength of gravity is decreased. The strengths of gravity in gravitation programs should be more optimized.

We have reported the results of the runs of the $g^* = 1.6$-$1.4$-$1.6$ program to demonstrate the validity of gravitational tempering, and we have written that the runs of the $g^* = 2.0$-$1.4$-$2.0$ program are in progress. In the future, we will perform simulations of other gravitation programs, which include complex programs such as $g^* = 1.6$-$1.4$-$2.0$, $g^* = 1.6$-$1.4$-$1.6$-$2.0$-$1.6$, and $g^* = 2.2$-$1.2$-$2.2$. Changes in the direction of the gravity and application of external share stress may work more effectively. In the present study, we fixed the template; however, tuning of the template pattern is promising. Recently, Hilhorst et al.[35] successfully controlled the defects at predetermined positions by using designed templates. Controlling the defects at desired positions are a step toward fabrication of optical circuits. Gravitational tempering in the colloidal epitaxy alone can potentially control the positions of the defects; if one can perform gravitational tempering with monitoring the motions the defects, defect structure can be controlled. At present, we can say that *in situ* observation of the defects in a centrifugation experiment is not an easy task.

The authors gratefully acknowledge the helpful comments from Dr. S. Matsuo.

**Supporting Information Available**
This information is available free of charge via the Internet at http://pubs.acs.org/

**Figure legends**

**Figure 1:** First example of the evolution of the center of gravity, the numbers of crystalline particles, and the snapshots. **a**, Evolution of the center of gravity, and numbers of fcc-like,

hcp-like, and other crystalline particles. **b-d**, the snapshots at the 2×, 2.2×, and 2.3×10$^7$th MCC [indicated by the arrows in **a**] in this order in course of the simulation. A significant sink of the center of gravity and change in the numbers of crystalline particles occurred during 2.25-2.3×10$^7$ MCCs. An expanded defect dramatically disappeared in this duration.

**Figure 2:** Second example of the evolution of the center of gravity, the numbers of crystalline particles, and the snapshots. **a**, Evolution of the center of gravity, and numbers of fcc-like, hcp-like, and other crystalline particles. **b-d**, the snapshots at the 2×, 2.5×, and 2.6×10$^7$th MCC [indicated by the arrows in **a**] in this order for simulation. Despite the lack of a sink of the center of gravity and change in the number of crystalline particles, the extent of a defect structure shrinks.

Description about Supporting Information

Filename: SI1.avi

Summary: An animation of the configuration of Figure 1 for the entire run.

Title and legend: Supplementary Video 1: Animation of the configuration of Figure 1. Disappearance of an extended triangular-shaped defect configuration occurred during 2.25-2.29×$10^7$ MCCs.

Filename: SI2.avi

Summary: An animation of the configuration of Figure 2 for 1.5-3×$10^7$ MCCs.

Title and legend: Supplementary Video 2: Animation of the configuration of Figure 2. Disappearance of an extended planner defect configuration occurred during 2.55-2.6×$10^7$ MCCs.

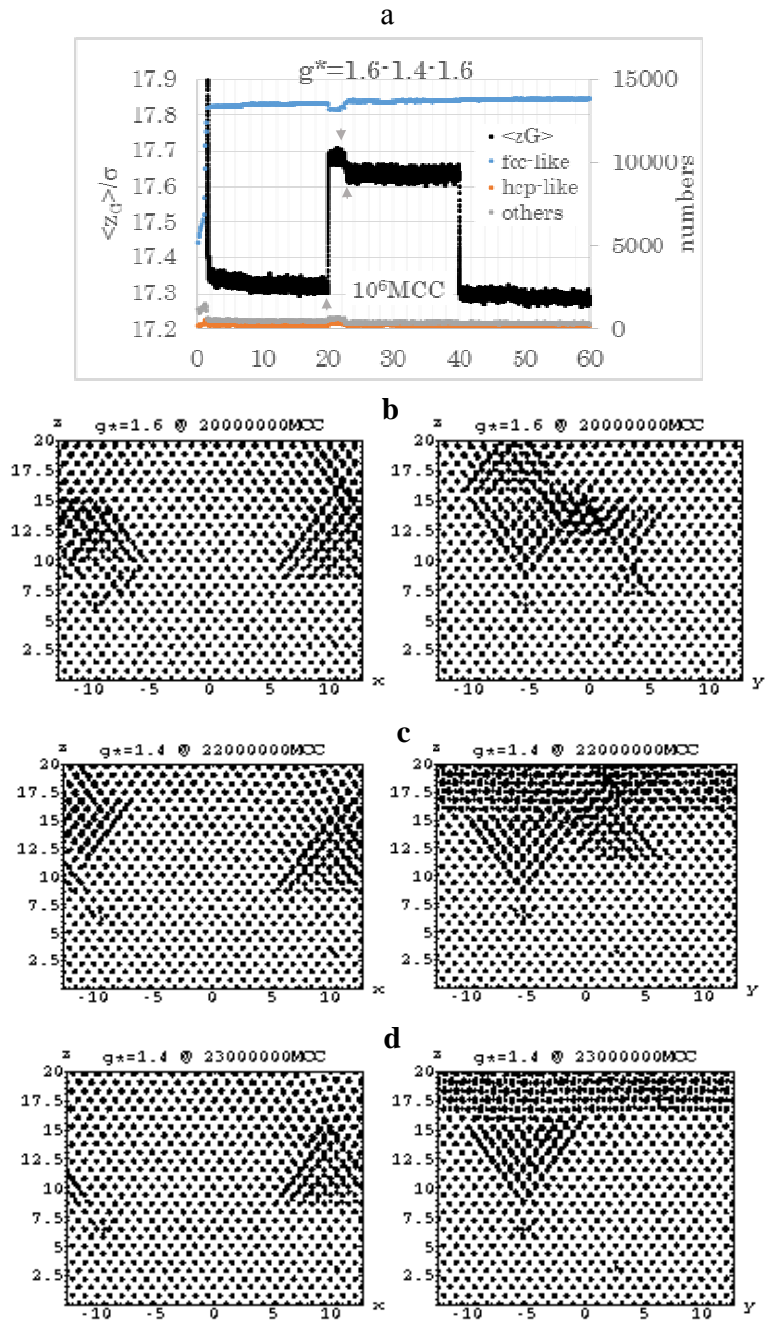

A. Mori et al.: Figure 1

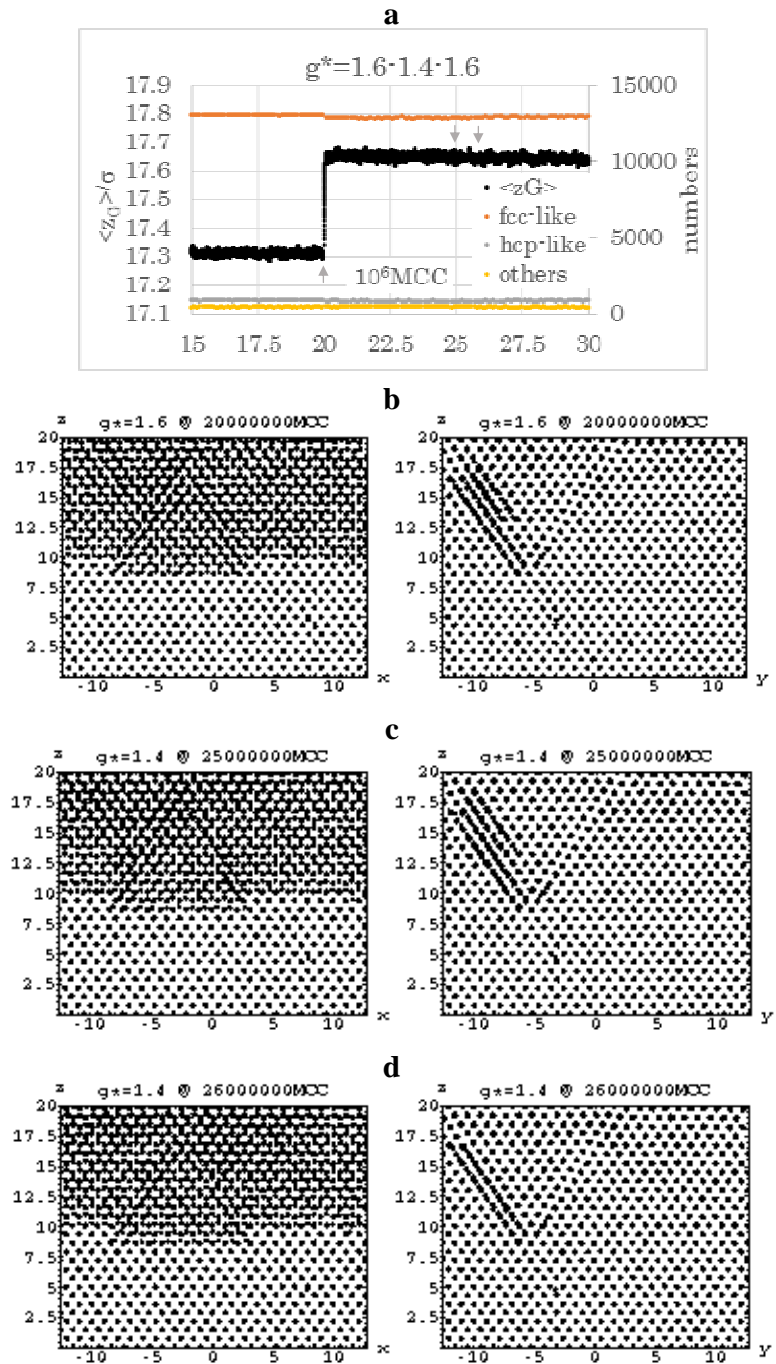

A. Mori et al.: Figure 2